\documentclass[11pt,a4paper]{article}
\usepackage{jheppub} 

\usepackage{amsmath,graphicx,amsfonts, mathrsfs, amssymb}

\newcommand{\be}{\begin{equation}}
\newcommand{\ee}{\end{equation}}
\newcommand{\beqa}{\begin{subequations}\begin{eqnarray}}
\newcommand{\eeqa}{\end{eqnarray}\end{subequations}}

\def\dd{\mathrm{d}}
\def\D{\Delta}
\def\tE {{\tilde E}}
\def\te {{\tilde e}}
\def\tB {{\tilde B}}
\def\tb {{\tilde b}}

\def\td{{ \tilde{n}_q }}

\def\r{{ u }}
\def\cS{{ \cal {S} }}
\def\cN{{ \cal {N} }}
\def\cE{{ \cal {E} }}
\def\tA{{\tilde A}}
\def\tF{{\tilde F}}
\def\tJ{{\tilde J}}
\def\tn{{\tilde n}}

\title{The open string membrane paradigm with external electromagnetic fields}

\author[a,d]{Keun-Young Kim,}
\author[b,d]{Jonathan P. Shock,}
\author[c,d]{and Javier Tarr\'{i}o}

\affiliation[a]{ School of Physics and Astronomy, University of
Southampton, \\ Southampton, SO17 1BJ, UK 
}
\affiliation[b]{Max-Planck-Institut f\"{u}r Physik (Werner-Heisenberg-Institut), \\
F\"{o}rhringer Ring 6, 80805 M\"{u}nchen, Germany.
}
\affiliation[c]{ Institute for Theoretical Physics, Universiteit Utrecht,\\
$\ \ $3584 CE, Utrecht, The Netherlands. 
}
\affiliation[d]{Kavli Institute for
Theoretical Physics China, CAS, \\ Beijing 100190, China.  
}

\emailAdd{k.kim@southampton.ac.uk}
\emailAdd{jonshock@mppmu.mpg.de}
\emailAdd{l.j.tarriobarreiro@uu.nl}


\subheader{SHEP-11-08, MPP-2011-36, ITP-UU-11/10}

\abstract{
We study the effective geometry felt by the fluctuations of open strings living on the worldvolume of probe D-branes in the presence of background electromagnetic fields. This is captured by an effective action consisting of a Maxwell term and a topological term, with the role of the metric played by the open string metric. Studying generalized Eddington-Finkelstein coordinates for stationary but non-static manifolds, we consider an open string membrane paradigm to obtain a generic formula for the DC transport coefficients, including the effect of external electromagnetic fields present on the worldvolume of the probe branes. We show that the previously studied singular shell, present when a critical electric field strength is turned on, behaves as a horizon for the open string degrees of freedom. The results of this analysis can be used to define a membrane paradigm for a very general class of spacetimes with non-diagonal metrics.
}

\begin{document}

\maketitle

\section{Introduction}

The most powerful results to come from the AdS/CFT correspondence, and thus arguably from string theory, are those of a universal nature. Although we are some way from finding the holographic dual of QCD, or indeed of any real-world condensed matter system, we have found that there are certain quantities which are invariant in a wide class of theories. It is clear that such universal quantities should not depend on the microscopic nature of the theory at hand and thus those which are of a low-energy nature or at critical points in the phase diagram are the most striking. Of these, one of the most celebrated is 
the ratio of the shear viscosity to entropy density of large $N$ gauge theories \cite{Kovtun:2003wp}. 

The ratio $\eta/s=1/4\pi$ was known well before the advent of the AdS/CFT correspondence \cite{Damour:1978cg}, and is a famed result of the black hole membrane paradigm which shall be discussed in the following section. The link to holographic gauge theories was shown in \cite{Kovtun:2003wp} by considering fluctuations on top of a supergravity background described by a Maxwell action. Using the prescription given in \cite{Son:2002sd}, it was possible to obtain the shear viscosity from the hydrodynamic expansion of the retarded correlator.

The connection between the membrane paradigm result and holographic transport coefficients was put on a solid footing in \cite{Iqbal:2008by}, where it was shown that at the level of linear response theory, the holographic properties of a strongly coupled thermal field theory are determined, in the low frequency and momentum limit, by the horizon geometry of the gravitational dual. With this understanding, the universality of the ratio $\eta/s$ has been restricted to gravitational theories described by isotropic Einstein gravity (plus a cosmological constant) \cite{Buchel:2008vz,Erdmenger:2010xm}.

However, the results presented in  \cite{Iqbal:2008by} are not always applicable. The assumptions considered in that work fail to capture the physics of probe branes with non-trivial field strength components on their worldvolumes. Non-zero field strengths are frequently needed to capture the phenomenology of the systems modelled by the AdS/CFT correspondence. For example, it is essential to turn on the temporal component of an abelian gauge field to describe, via the holographic dictionary \cite{Gubser:1998bc,Witten:1998qj}, a chemical potential. It is also interesting to include the effects of external electromagnetic fields in the models describing both the Quark-Gluon Plasma and lower-dimensional condensed matter systems.

In \cite{Mas:2008qs} some of us studied the generalization of the formula for the DC conductivity provided in \cite{Iqbal:2008by}, applied to probe $Dp$-branes at finite baryon density. The main difference between both results is that in  \cite{Mas:2008qs} the r\^ole of the metric is played by the non-symmetric quantity $\gamma_{mn}= g_{mn}+2\pi\alpha' F_{mn}$, where $g_{mn}$ is the pullback of the $10$-dimensional metric onto the worldvolume of the probe branes, and $F_{mn}$ is the field strength associated to the $U(1)$ gauge field living in the probe branes. The formula for the DC conductivity given in  \cite{Mas:2008qs} recovers previous results from the vanishing electric field limit of the result in \cite{Karch:2007pd}, where the DC conductivity is found by demanding reality of the action, and hence no fluctuations are involved. We will refer to this last calculation as the \emph{macroscopic} one, whereas that in which fluctuations are studied will be referred to as the \emph{microscopic} result.

The microscopic calculation in the case of an external electromagnetic field was studied in \cite{Mas:2009wf}. It was shown that with the imposition of the appropriate boundary conditions at a special position on the probe-brane worldvolume (the `singular shell') was enough to recover the macroscopic results for the conductivity via a Kubo relation. The Kubo relation is understood to give the conductivity in linear response theory. In the case of finite background fields, the response to an infinitesimal electric field must be interpreted as an infinitesimal addition to a finite background value and thus the microscopic and macroscopic calculations agree, even when one is beyond the linear response regime.

In the present work we show that the microscopic determination of the DC transport coefficients performed in  \cite{Iqbal:2008by} can be extended to a new form of membrane, associated with the interactions of open string states in an asymptotically $AdS$ background. Through the use of the open-string metric it is possible to show that the degrees of freedom on a probe $Dq$-brane feel an induced horizon due to the introduction of certain background fields (in particular those corresponding to turning on an external electric field in the gauge theory). We are able to rederive holographic results in the language of a new membrane paradigm.

We will also discuss the nature of the induced horizon and propose that its temperature should be read as the temperature of the gauge theory felt by fundamental matter moving in the electric field. Thus, the results of this paper in summary are:
\begin{itemize}
\item The geometrization of background gauge fields on probe branes leads to an effective action for open string degrees of freedom via the open string metric which may include both a Maxwell term and a topological term.
\item Using open string degrees of freedom we find that an electric field on a probe brane induces a horizon structure with an associated temperature which can be interpreted in the dual gauge theory language. We extend this to the case which includes finite baryon density and background temperature.
\item The open string horizon can be interpreted in membrane paradigm language, leading to the calculation of transport coefficients on the electric membrane.
\item Quasinormal modes in the open string metric also see the horizon and we can thus define retarded and advanced Green's functions in the usual way via appropriate boundary conditions on the induced horizon with purely incoming boundary conditions.
\end{itemize}

 \paragraph{The black hole membrane paradigm}

Before studying the gravity dual of flavour systems in the presence of background fields, let us summarize the classical membrane paradigm following closely the exposition in \cite{Iqbal:2008by}. 
Here we review the basic idea with only a Maxwell term in the \emph{diagonal} metric background. 
We will extend the method to a Maxwell plus topological term in \emph{non-diagonal} and
 \emph{non-static, stationary} metrics in sections \ref{sec.mempar1} and \ref{sec.mempar2}.  

The membrane paradigm and its link to holography is a consequence of a few very simple results. 
In defining the action for a spacetime with a horizon we are forced, by the consistency of the variational principle, to include a surface term on the horizon itself. One can ask what an observer sees when hovering just above this surface, by looking at how terms in the bulk show up as surface terms on the horizon. One considers not the action on the horizon itself as the surface term, 
but on a slice at radius $r_0$ just outside the horizon, the so called \emph{stretched horizon}. 
For example, consider a Maxwell term in the bulk,
\be
S_{\mathrm{bulk}}=-\int_{r>r_0}\!\!\!\!\!\!\!\!\!\dd r\,\dd^dx\sqrt{-g}\,\frac{f_{MN}f^{MN}}{4\,g_{d+1}^2(r)} \, ,
\ee
where $M,N$ run over all spacetime directions in the bulk, $g_{d+1}(r)$ is an $r$ dependent coupling term and $g$ is the determinant of the metric in the bulk. This action induces a surface term
\be
S_{\mathrm{surf}}=\int_\Sigma\sqrt{-\gamma}\left(\frac{{\cal J}^\mu}{\sqrt{-\gamma}}\right)a_\mu \, ,
\ee 
to cancel the boundary term arising from the variation of the bulk Maxwell action. 
$\gamma$ is the induced metric on the spatial slice at radius $r_0$ and 
\be
{\cal J}^\mu=-\frac{\sqrt{-g}\,f^{r\mu}}{g_{d+1}(r_0)^2} \, .
\ee
Such a surface term looks to an observer hovering at the slice $r_0$ to be a current, $\frac{{\cal J}^\mu(r_0)}{\sqrt{-\gamma}}$, induced by the source $a_\mu$. By demanding a regular field strength on the horizon (given that for an infalling observer there is nothing special about this radius), the dependence on the combination of $r$ and $t$ is constrained, and one can thus show that, for a static spacetime, $f_{ri}=\sqrt{\frac{g_{rr}}{g_{tt}}}f_{ti}$, thus linking the current and electric field on the horizon. The Maxwell equations, together with the Bianchi identities, lead to a current
\be
j_{mb}^i \equiv \frac{{\cal J}^i(r_0)}{\sqrt{-\gamma}} =\frac{1}{g_{d+1}^2}\sqrt{\frac{g}{g_{rr}g_{tt}}}g^{ii}\Bigg|_{r_0} f_{it} \, ,
\ee
where $f_{it}$ is the electric field strength measured by the observer at the horizon. Such a response current to an electric field leads to the interpretation of the horizon as a conducting membrane with conductivity
\be
\sigma_{mb} \equiv \frac{1}{g_{d+1}^2}\sqrt{\frac{g}{g_{rr}g_{tt}}}g^{ii}\Bigg|_{r_0}\, .
\ee

The key point to the link between the membrane paradigm and the AdS/CFT correspondence was the fact that, in the hydrodynamic limit, the expression for the current on the boundary of $AdS$ is made up of terms which together are independent of the radial coordinate. Thus, the expression can be taken from the boundary to the horizon, and is found to match exactly with the conductivity of the membrane. The non-trivial evolution in the radial direction takes into account effects beyond the hydrodynamic limit.

 \paragraph{Outline of this paper} 
The discussion we are concerned with requires some intuition on the behaviour of probe branes in the presence of external fields. In section \ref{sec.metallic} we review the most important aspects of this setup, providing the references needed to study the published work in further detail. We also provide some expectations from the field theory in order to justify the validity of the probe approximation which we use throughout.
 
 The reader familiar with such systems will not find anything new in section \ref{sec.metallic}, and may skip directly to section \ref{sec.flucs}, where we discuss the main ideas contained in this paper. First we write an effective action for the gauge field fluctuations on the probe brane, in terms of a metric capturing the effects of the background electromagnetic fields in the bulk. We then revisit the membrane paradigm assuming the effective metric is static and show in an example how we can easily study DC transport coefficients.
 
When we have an electric field and finite baryon density in the background, the effective metric felt by the vector fluctuations is not static but stationary (in the probe approximation). We study the properties of the horizon of this effective metric in section \ref{sec.sshell}, and extend the discussion on the membrane paradigm to non-static, stationary, non-diagonal metrics, giving examples of the application of our results.

We conclude in section \ref{sec.conclu} by commenting again on the most important results in this paper, together with some outlook for future work.

 % END OF INTRODUCTION %

\section{Metallic AdS/CFT}\label{sec.metallic}

The study of gauge theories in the presence of electromagnetic fields in a holographic context has, over the last few years, been under intense investigation \cite{Filev:2007gb,Karch:2007pd,Filev:2007qu,O'Bannon:2007in,Karch:2008uy,Ammon:2009jt,Albash:2007bk,Erdmenger:2007bn,Albash:2007bq,Bergman:2008sg,Kundu1,Kim:2008zn,Filev:2008wj,Kundu2,Bergman:2008qv,Filev:2009xp,Mas:2009wf,Evans:2010xs,Evans:2010tf}. As is frequently the case in the holographic correspondence, properties in the gauge theory are mapped beautifully to geometric properties of the gravity picture. We shall first go through in detail the physics of the gauge theory description before going through the parallel analysis in the gravity picture.

\subsection{Gauge theories in the presence of background electric fields}
The construction of interest is given by applying an external electric field, $E_x$, in the $x$-direction in ${\mathbf M}^{1,3}$ to $N_f$ hypermultiplets of ${\cal N}=2$ charged fundamental matter interacting with ${\cal N}=4$ super Yang-Mills with gauge group $SU(N_c)$ at finite temperature. The ratio $N_f/N_c$ is taken to vanish such that, while we can consider processes involving the propagation of fundamental matter, fundamental loops are suppressed. For low temperatures and low electric field strengths ($T/m_q\lesssim 1$,$\sqrt{E_x}/m_q\lesssim 1$) quarks are bound into mesons which are infinitely stable (in the 't Hooft limit), with a discrete, and gapped spectrum. As the temperature or electric field strength is increased, the mesons are destabilized as the electric field pulls the charged particles apart. The thermal fluctuations destabilize them further. The effect of the electric field will thus be to pair produce if its energy is high enough with respect to the mass of the fundamental matter.

As one turns up the electric field to a value such that pair production initiates, the charged quarks and antiquarks\footnote{We will use the word quarks generally when talking about charged fundamental matter.} which are pulled out of the vacuum are accelerated in opposite directions and quickly collide with the background of adjoint matter, whose energy density is $\mathcal O\left(N_c/N_f\right)$  higher than that of the fundamental matter. The adjoint matter thus acts as a bath, which can absorb, in the quenched approximation, an infinite amount of energy from dissociated mesons without heating up. On colliding with the adjoint matter, some of the velocity is deposited in the background, thus heating it (infinitesimally) but not adding  overall momentum, because for every quark moving in one direction there will be an antiquark moving in the other. 

Clearly, this flow of charge corresponds to a current, which, on switching on an electric field at time $t_0$, will be time-dependent. In the quenched approximation this time dependence will show up only at short times and asymptotically long times, parametrically of order $\mathcal O\left(N_c/N_f\right)$, while the charge is accelerated until it comes to a pseudo-equilibrium with the adjoint matter, such that the energy flow from the electric field equals the energy flow from the fundamental into the adjoint matter. In the present work we will take the perspective that we have already come to this pseudo-equilibrium and thus a semi-steady state has been established (see \cite{Karch:2010kt} for a time-dependent solution to such a process). We can thus talk about the conductivity of the plasma, given by the ratio of the induced current to the electric field strength, $\sigma=\frac{J_x}{E_x}$.

As the fundamental matter moves through the background of adjoint matter at finite temperature, it feels a hot wind from the oppositely charged fundamental matter coming from the other direction which have also come from pair production process. This temperature is hotter than the temperature of the bath and is a combination of the relativistic effects of a velocity on the energy density of the background (see \cite{Liu:2006nn}) and also of the thermal interactions with other fundamental matter. We denote the effective temperature $T_\mathrm{eff}=T_\mathrm{eff}(T,E_x,m_q)$. Indeed, if the background is at zero temperature, the accelerated fundamental matter will still feel an effective temperature -- it will appear to be in a bath of temperature $T_\mathrm{eff}$. If one includes the effect of fundamental loops, the adjoint matter would heat up as more and more energy from the electric field was transferred to thermal energy through collisions with the fundamental matter. There is no equilibrium in this case as the plasma will heat up indefinitely and thus beyond the probe approximation the conductivity is infinite -- as expected for a translationally invariant system.

In the presence of a finite baryon density the situation is similar, but in this case one does not need to reach a critical electric field value or temperature to set up a current. The matter making up the quark density will be accelerated, adding to the current induced by pair-production at any finite electric field value. In this case not only is there a finite current, but also a finite momentum flow. The effective temperature will thus also be a function of the baryon density. On going beyond the quenched approximation in this case, not only will the system heat up indefinitely, but it will also accelerate in a direction determined by the sign of the baryon density.

Thus we have a simple picture of this pseudo-equilibrium situation of charged fundamental matter interacting with a large number of ${\cal N}=4$ hypermultiplets at finite temperature in the presence of an electric field with, or without finite baryon density. Such a simple picture is perfectly mirrored in its holographic dual.

\subsection{D-branes in the presence of background electric fields}\label{sec.fieldtheory}
The gravity dual of the above setup is very familiar and has been studied in detail in a large number of papers (see previous references in this section). However, here we would like to reexamine this picture and both reinterpret the known results and derive new ones in the language of the membrane paradigm.  We will focus in detail on the $D3/D7$ brane intersection in the quenched approximation, though many of the ideas are paralleled in similar intersecting models.

The supergravity geometry that we are interested in is the $AdS_5 \times S^5$ background generated by $N_c\gg 1$ black $D3$-branes, that may be parametrized as 
\begin{eqnarray}
  \dd s^2 &=& G_{MN} \dd x^M \dd x^N   \\
  &\equiv& \frac{(\pi T L)^2}{\r}\left(-f \dd t^2+\dd \vec{x}^2\right)+ \frac{L^2}{4f\r^2} \dd\r^2
   + L^2 \left( \dd \theta^2 + \sin^2\theta \dd\Omega_3^2 + \cos^2\theta \dd\phi^2 \right) \nonumber \ ,
\end{eqnarray}
where $L$ is the radius of the $AdS_5$ and $S^5$ spaces, $\dd \vec{x}^2 \equiv \dd x^2 + \dd y^2 + \dd z^2$, 
$\dd \Omega_3$ and  $\dd \phi$ are the metrics of unit radius $3$ and $1$-spheres respectively. The blackening factor is $f\equiv1-\r^2$, where $\r$ is dimensionless\footnote{The dimensionless radial coordinate $u$ is related to 
the standard AdS radial coordinate, $r$ in  
$ \dd s^2 = \frac{r^2}{L^2}\left(-f \dd t^2+\dd \vec{x}^2\right)+ \frac{L^2}{r^2} \frac{\dd r^2}{f}
   + L^2 \dd \Omega_5^2 \ , \  \left(f(r) = 1-\frac{r^4_H}{r^4}\right) $, by $u \equiv \frac{r_H^2}{r^2} \equiv \frac{(\pi T L^2)^2}{r^2}$. 
} and 
goes from the boundary at 0 to the horizon at $\r=1$.

It is in this background that we will study the phenomenology of $D7$ probe branes. The DBI action for $N_f\ll N_c$ $D7$-branes is written in terms of the pullback metric from the bulk within which it lives, plus the scalar, fermionic and vector valued fields on its worldvolume. In the following we will turn off all fermionic terms
\be\label{eq.dbi}
S_{DBI} = -T_{D7} N_f \int \dd^{8}\xi\, e^{-\phi} \sqrt{-\det(\gamma_{mn})} \, .
\ee
where $T_{D7} = (2\pi)^{-7}\alpha'^{-4}$ and $\gamma$ denote the pullback metric plus gauge field\footnote{Our index conventions
are the following: (1) background spacetime: $MNPQ=0,1,\cdots,9$; (2) worldvolume spacetime: $mnpq = 0,1,\cdots, 7$; (3) 4D field theory spactime: $\mu\nu\rho\sigma = 0,1,2,3$; (4) 3D field theory space: $i,j,k=1,2,3$. 
Also note that the gauge field with tilde includes the factor of $2\pi\alpha'$, i.e. $\tilde{A}_M = 2\pi\alpha' A_M$. },
\be\label{eq.gamma}
\gamma_{mn} \equiv \partial_m X^M \partial_n X^N G_{MN}+ 2\pi\alpha' F_{mn} \equiv g_{mn}+\tF_{mn} \, .
\ee
The $S^3$ direction is trivial so it is convenient to integrate out and the action becomes 5 dimensional.

We chose the embedding with $\theta=\theta(u)$ and $\phi=0$.
The pullback metric of a $D7$-brane onto the $D3$-brane background will thus be given by
\begin{eqnarray}\label{inducedmetric}
&& \dd s_{7+1}^2 = g_{tt} \dd t^2 + g_{xx} \dd \vec{x}^2 + g_{\r\r}\dd \r^2 + g_{\Omega\Omega}^2 \dd\Omega_3^2   \label{gab}  \\
&& \qquad \equiv \frac{(\pi T L)^2}{\r}\left(-f \dd t^2+\dd \vec{x}^2\right)+L^2 \left(\frac{1}{4f\r^2}+\frac{\psi'^2}{1-\psi^2} \right) \dd\r^2+L^2 \left(1-\psi^2\right)\dd\Omega_3^2 \ , \nonumber 
\end{eqnarray}
where $\psi(\r) \equiv \cos\theta(\r)$ is the field which defines the embedding of the $D7$-brane in the black $D3$-brane geometry.

We will consider different choices for the specific gauge field on the $D7$-brane, but the one-form gauge potential ($A$, such that  in equation \eqref{eq.gamma} $F=dA$) which captures all phenomena discussed here is
\begin{eqnarray}
  2 \pi \alpha' {A} &\equiv&  \tA + 2 \pi \alpha' a  \nonumber \\ 
  &\equiv& \tA_t(\r) \mathrm{d} t  + (- \tE_x t + \tA_x(\r) ) \mathrm{d} x  + (\tB_z x+\tA_y(\r)) \, \dd y \nonumber \\
  & &
  + (\tB_x y+\tA_z(\r)) \, \dd z   + 2 \pi \alpha' a   \ , \label{AA}
\end{eqnarray}
where tilde variables include $2\pi \alpha'$ and correspond to the background field. $A_t$  introduces finite baryon density, $E$ is a constant background electric field and $B$ a magnetic field. $A_x$, $A_y$ and $A_z$ will encode the optical and Hall currents generated by the electric field and magnetic fields in the presence of baryon density. On top of this, we will consider fluctuations $a_{x,y,z}(t,\r)$ in section \ref{sec.effaction}.  

In the presence of the electric field, if the field is large enough as compared to the mass of the fundamental matter, there is a special position on a probe $D7$-brane's worldvolume at which its Lagrangian density develops a complex part in the absence of a current, and vanishes in its presence. This radius defines a shell around the origin of $AdS$ and shall be termed the \emph{singular shell}. The study of the singular shell and its interpretation as a horizon will make up a large part of this paper.

\subsection{The induced current and the singular shell}

In this section we simplify to the case of zero magnetic field and show how the singular shell appears naturally in the setup. 
In this case the DBI action at the classical level (\emph{i.e.}, turning off fluctuations) is given by %
\begin{eqnarray}
  && \cS = - \cN \int \dd t \,\dd\vec{x} \,\dd u\, g_{xx} g_{\Omega\Omega}^{3/2} \sqrt{-g_{uu}(g_{tt}g_{xx} + \tE_x^2) - g_{xx} \tA_t'^2 - g_{tt} \tA_x'^2 }    \ , \\
  &&  \cN \equiv N_f T_{D7} g_s^{-1} 2\pi^2 = \frac{\lambda N_f N_c}{(2\pi)^4 L^8}     \ ,
\end{eqnarray}
where $2\pi^2$ is a volume factor of $S^3$ and $L^4 \alpha'^{-2} = 4\pi g_s N_c = g_{YM}^2 N_c = \lambda$. 

Since the gauge fields enter the action only through their field strengths there are conserved charges
\begin{equation}
\hat{n}_q \equiv \frac{1}{\cal N}\frac{\delta  {\cal S}}{\delta \tA_t'(\r)}\, , 
\,\,\,\,\, \hat{J}_x \equiv \frac{1}{\cal N}\frac{\delta  {\cal S}}{\delta \tA_x'(\r)}    \ ,
\end{equation}
where $\hat{n}_q$ is related to the quark density \cite{Kobayashi:2006sb} and $\hat{j}_x$ is related to the current in the $x$-direction \cite{Karch:2007pd}. A normalization $\cal N$ is chosen to simplify the Legendre transformed action and a further, physical normalization is shown later in equations (\ref{dimless}) and (\ref{trans1}). 
The Legendre transformed action reads
\begin{eqnarray}
  \cS_{LT} &=&  \cS - \int \dd t \,\dd \vec{x}\, \dd u \left[
  \tA_t' \frac{\delta S}{\delta \tA_{t}'} + \tA_x' \frac{\delta S}{\delta \tA_{x}'} \right]    \ , \nonumber \\
   &=& - \cN \int \dd t \,\dd \vec{x} \,\dd u   \sqrt{ - (g_{tt} g_{xx} + \tE_x^2)g_{xx}^2 g_{\r\r} g_{\Omega\Omega}^3 }
  \sqrt{1+ \frac{g_{tt} \hat{n}_b^2 + g_{xx} \hat{J}_x^2}{g_{tt} g_{xx}^3 g_{\Omega\Omega}^3}} \, .
\end{eqnarray}
Since $g_{tt}g_{xx} < 0$, the first square root changes sign at a radius $\r_s$ (hereby called the singular shell) such that  
\be
\tE_x=\sqrt{-g_{tt}(\r_s)g_{xx}(\r_s)}\, ,\quad \r_s=\frac{1}{\sqrt{1+\tilde{e}_x^2}} \, ,
\ee
where $\tilde{e}_x=\frac{\tE_x}{(\pi T L)^2}$. 
In order to make the action always real this sign change should be 
countered by a sign change in the second term, which implies the relation:

\begin{eqnarray}
\tJ_x= \sqrt{\sqrt{1+\tilde{e}_x^2}\left(1-\psi_s^2\right)^3+\frac{\tn_q^2}{1+\tilde{e}_x^2}}\,\tilde{e}_x \label{Jx} \, ,
\end{eqnarray}
where 
\begin{eqnarray}
  \tn_q= \frac{\hat{n}_q}{(\pi T L^2)^3} \ , \quad \tJ_x=\frac{\hat{J}_x}{(\pi T L^2)^3} \ , \quad \psi_s = \psi(u_s) \ .
  \label{dimless}
\end{eqnarray}
These are convenient dimensionless variables which will be used in the relevant computations. 
We will interpret the final results in terms of dimensionful physical quantities 
$n_q$ (quark density), $J_x$ (current density), $E_x$ (electric field) which are related to 
their dimensionless counterparts as \cite{Kobayashi:2006sb, Karch:2007pd} 
\begin{eqnarray}
  \tn_q = \frac{8}{N_f N_c \sqrt{\lambda} T^3} n_q \ , \quad \tJ_x = \frac{8}{N_f N_c \sqrt{\lambda} T^3} J_x \ ,
  \quad  \te_x = \frac{2}{\pi \sqrt{\lambda} T^2 }E_x \ . \label{trans1}
\end{eqnarray}

A useful expression in terms of physical variables is 
\begin{eqnarray}
J_x = \sigma_{xx} E_x &=&   \frac{N_f N_c T}{4\pi}\sqrt{\sqrt{1+\tilde{e}_x^2}\left(1-\psi_s^2\right)^3+\frac{\tn_q^2}{1+\tilde{e}_x^2}}\, \ E_x \label{Jx1} \, , \\
&=& \sqrt{ \frac{N_f^2 N_c^2 T^2}{16 \pi^2} \sqrt{1+\tilde{e}_x^2}\left(1-\psi_s^2\right)^3+\frac{d_q^2}{1+\tilde{e}_x^2}}\, \ E_x\nonumber
\label{Jx2}
\end{eqnarray}
where $\sigma_{xx}$ is a conductivity. Note that $d_q \equiv \frac{N_f N_c T \tn_q}{4\pi} = \frac{n_q}{(\pi/2)\sqrt{\lambda}T^2}$ has dimension one, while $\te_x = \frac{E_x}{(\pi/2)\sqrt{\lambda}T^2}$ is dimensionless. The latter agrees with (3.7) in \cite{Karch:2007pd}, where $D \rightarrow n_q$ 
and $ B \rightarrow J_x$.

Thus, given a value for the embedding position, $\psi_s$, at the singular shell (or, by integrating the embedding to the boundary, the value of the mass of the fundamental matter), the baryon density and the electric field strength, we find a fixed current, as expected from the physics of the problem
As discussed in section \ref{sec.fieldtheory}, for zero baryon density and large mass fundamental matter ($\psi_s=1$) there is no current as the electric field is not large enough to pair produce\footnote{ We are abusing notation here, with  $\psi_s=1$ we refer to the  maximum value of $\psi$ taken as the brane approaches the $D3$-branes (though it may never pierce the singular shell). For Minkowski embeddings the brane solution reaches $\psi=1$, which is the boundary point from which the numerical integration is taken. The important point to note is that for very massive embeddings, the pair-creation term vanishes.}.

\begin{figure}[]
\centering
  {\includegraphics[scale=0.6]{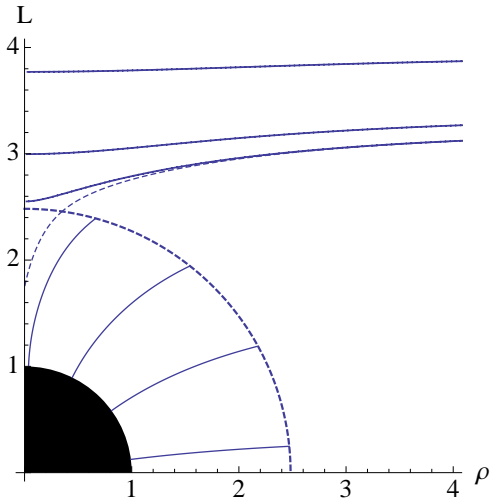}}
  \caption{\em \label{fig:D7embeddings} $D7$-brane embeddings in a the black $D3$-brane background with worldvolume electric field. The $L$ and $\rho$ coordinates, which are easier for visualising the embeddings, label a cartesian version of the $\r$ and $psi$ coordinates used throughout this paper. The black disk shows the black hole, while the circular dashed line labels the singular shell. Those embeddings which pass through the singular shell are in the conducting phase while those which do not are in the insulating phase. The dashed embedding exhibits a conical singularity on the $L$-axis.}\label{fig.embs}
  \end{figure}

In figure \ref{fig.embs} we show several representative embedding profiles, corresponding to different masses of the fundamental mass, at fixed electric field and zero baryon density. The black disk corresponds to the black hole, while the dashed circle defines the singular shell. Those embeddings which pierce the singular shell correspond to conducting fundamental matter, while those which do not have fundamental matter in the insulating phase. The dashed embedding which pierces the singular shell and then hits the $L$-axis has a conical singularity. However, as we will show, from the point of view of the fluctuations of open strings, the physics inside the singular shell is hidden from them and thus the conical singularity  is not a problem. According to \cite{Das:2010yw}, the conical singularity is a signature that the probe brane approximation needs a sink for the energy from the electric field to flow to.

This analysis have been done for a general case including a magnetic field (\ref{AA}) in \cite{Ammon:2009jt} and
we present the final conductivity results\footnote{ 
The results are valid for any black hole embedding. However, there may be
more than one embedding for a given set of parameters and the embedding of the lowest free energy will be
the ground state. To identify the ground state, we need to study the phase diagram \cite{Evans1}
following, for example, \cite{ Kobayashi:2006sb, Evans2,Evans3,Evans4}. }. 
\beqa
  \sigma^{xx} &=& \frac{N_f N_c T}{4 \pi }\sqrt{\left(1-\psi_s^2\right)^3  + \frac{\tn_q^2}{ \sqrt{{\cal F}}\left({\cal F}+\tb_x^2+\tb_z^2\right)}    }\frac{{\cal F}+\tb_x^2}{\sqrt{\sqrt{{\cal F}}\left({\cal F}+\tb_x^2+\tb_z^2\right)}} \ , \label{sxx0} \\
   \sigma^{yx} &=& -\frac{N_fN_c T}{4\pi} \frac{\tn_q \tb_z}{{\cal F}+\tb_x^2+\tb_z^2}   \ , 
   \qquad \sigma^{zx} = \frac{\tb_x \tb_z}{ {\cal F}+ \tb_x^2} \sigma^{xx} \ , \label{szx0}
\eeqa
where 
\begin{eqnarray}
  \tb_{x,z} &\equiv& \frac{B_{x,z}}{(\pi/2) \sqrt{\lambda} T^2} \ , \\
  {\cal F} &\equiv& \frac{1}{2} \left(1 + \te_x^2 - \tb_x^2 -\tb_z^2  + \sqrt{(\te_x^2 - \tb_z^2)^2  + (1+\tb_x^2)(1+\tb_x^2 + 2(\te_x^2 + \tb_z^2))}\right) \ ,
\end{eqnarray}
and $\rho$  in (3.7) of \cite{Ammon:2009jt} is the same as our $d_q = \frac{N_f N_c T \tn_q}{4 \pi } = \frac{n_q}{(\pi/2)\sqrt{\lambda}T^2}$ defined below (\ref{Jx1}).
Note that the limit $\te_x \rightarrow 0$ implies  ${\cal F} \rightarrow 1$.

\subsection{Thermodynamics of the probe-brane system}
The evidence in this paper will suggest that the fluctuations on the probe-brane worldvolume feel the singular shell as a horizon. In order to study the thermodynamics of the system one needs to integrate over the worldvolume of the probe-brane. It appears that in order to calculate the free energy one should include the region inside the singular shell and thus thermodynamic quantities are concerned with the behaviour inside this region. However, the DBI action itself breaks down between the singular shell and the background horizon.

The DBI action is an approximation in the slow-varying limit of $F$ on the brane worldvolume (slow such that it doesn't vary on the string scale). In the presence of a finite electric field, when a current is set up, this limit is no longer valid. While $A_x(\r)$ is well-behaved in and around the singular shell, it is singular at the background horizon
\begin{equation}
A_x(\r)\sim E_x \, \log\left( \r - \r_h \right) \ ,
\end{equation}
and thus the DBI action is not a good approximation in this region. In fact it appears that this may be a relic of the conical solutions, seen for embeddings which do not hit the background horizon but are Minkowski embeddings with conical singularities.

The breakdown appears to come from the lack of back-reaction in these systems. We are pumping energy into the brane through the electric field and in the probe approximation there is no sink for this energy. While the solutions outside the singular shell are well behaved and thus the transport coefficients can be trusted as these only depend on the behaviour outside this region, those inside the singular shell and close to the background horizon should not be trusted. Therefore, to perform a free energy calculation one has to understand better how to treat the action in this approximation.

%%% END OF SECTION 1

\section{The open string metric and fluctuations in external fields}\label{sec.flucs}

When background Kalb-Ramond fields or world-volume gauge fields on a probe D-brane are turned on, the fluctuations of open strings on the probe-brane do not feel simply the background geometry that they are probing \cite{Seiberg:1999vs, Gibbons:2000xe}. The open string metric describes precisely the effective geometry felt by open strings in the presence of external fields. Following \cite{Seiberg:1999vs}, we define the (inverse of) the open string metric, $s^{mn}$, and an antisymmetric tensor $\theta^{mn}$ through the simple relation
\be\label{eq.gamma}
\gamma^{mn}=\left(\gamma_{mn}\right)^{-1}=s^{mn}+ \theta^{mn} \, .
\ee
$s_{mn}$ is defined such that $s_{mn} s^{np}=\delta_m^p$, then
\beqa
s_{mn} &= &g_{mn}-(\tF g^{-1} \tF)_{mn} \label{eq.sab}  \, , 
\eeqa
We will refer hereafter to $s_{mn}$ as the open string metric (OSM). This object determines the equations of motion and thus solutions of the excitations of the probe brane\footnote{The fields living in the brane correspond to open strings degrees of freedom, hence the name OSM.}. 

To see this let us recall the blackfold dynamics of \cite{Emparan:2009at}. The profile of a brane embedded in a higher-dimensional manifold via some relations $f^a=0$, $a=1,\cdots,N$, is described by\footnote{We refer to \cite{Emparan:2009at,Grignani:2011mr} for details.}
\be\label{eq.blackfold}
K_{mn}^a T^{mn} = 0 \, , \qquad D_m T^{mn}=0 \, ,
\ee
where $K_{mn}^a$ is the extrinsic curvature with $m,n$ along the $Dq$-brane worldvolume, $D_m$ is a covariant derivative with respect to the induced metric and $T^{mn}$ the stress-energy tensor of the membrane. The first relation in \eqref{eq.blackfold} is a generalization of the geodesic equation which makes it clear that the embedded $D7$-branes have an extremal volume. The fact that $T_{mn}$, and not the metric, enters into equation \eqref{eq.blackfold} is a manifestation  of the brane's profile being determined not only by gravitational interactions in the $10$-dimensional bulk, but also by the field strength on the brane's worldvolume.

 In the case at hand, the probe branes are described by the DBI action \eqref{eq.dbi}, and the stress energy tensor is found to be \cite{Grignani:2011mr}
\be\label{eq.tmunu}
T^{mn} = \frac{2}{\sqrt{-g}} \frac{\delta S_{DBI}}{\delta g_{mn}} = -T_{D7} N_f  g_{d+1}^2 s^{mn}\, ,
\ee
where we have defined a function $g^2_{\mathrm{d+1}} \equiv \sqrt{-\gamma}/\sqrt{-g}= \sqrt{-s}/\sqrt{-\gamma}$, which induces a conformal transformation in $s_{mn}$. The equations of motion derived from \eqref{eq.blackfold} coincide with the ones obtained via the Euler-Lagrange equations from the action \eqref{eq.dbi}, showing that $s_{mn}$ and $g_{d+1}^2$ are all we need to determine the profile of the fields on the brane.

A similar relation allows one to express $\theta^{mn}$ as the variation of the DBI action with respect to the field strength $F_{mn}$. This means that $s$ and $\theta$ are conjugate variables to $g$ and $F$, and therefore we can make a Legendre transformation of the action to express it in terms of the new variables. Such a procedure will give rise to a topological term coming from
\be
 \frac{\delta S_{DBI}}{\delta F_{mn}} F_{mn}= N_f T_7 \int \dd^8 \xi \sqrt{-\gamma}\, \theta^{mn}F_{mn}   \ ,
\ee
which can be written as a total derivative and therefore contributes only with a boundary term. In this paper we will not pursue this direction further, though the further study of this term may be necessary to understand the thermodynamics of some brane-intersection scenarios.

\subsection{The effective action for fluctuations of gauge fields on probe branes}\label{sec.effaction}

Consider now the fluctuations of the abelian gauge field living on a probe brane\footnote{Note that in the following we do not consider the Wess-Zumino term which will not be important for the modes discussed here and in general will not alter the qualitative form of the effective action derived.}. We will not consider the fluctuations along the $S^3$ wrapped by the $D7$-branes (thus by expanding the DBI action to second order in gauge field fluctuations we can integrate over this $S^3$). Considering the quadratic fluctuation terms we are left with an effective  action which can be written in terms of $s_{ab}$ as
\beqa
  S_{\mathrm{eff}} & = & - \mathcal{N}' \int \dd^{5}x 
  \left[\frac{ \sqrt{-s} }{4\, g^2_{5}} s^{mp}s^{nq} f_{mn} f_{pq} 
   + \frac{1}{8} \epsilon^{mnpqr} f_{mn}  f_{pq}Q_r  \right] \label{eq.generalmeson}\, , \\
  \mathcal{N}' & \equiv & {\cal N} (2\pi\alpha')^2 = \frac{N_f N_c}{4\pi^2 L^4} \, , \qquad 
    Q_r = -\frac{\sqrt{-\gamma}}{8} \epsilon_{mnpqr}\theta^{mn}\theta^{pq} \, , \label{eq.gm2}
\eeqa
where $g^2_{5} = \sqrt{-s}/\sqrt{-\gamma}$ and the indices do not run along the $S^3$ directions (notice however that $\sqrt{-s}$ knows about the components of the metric along these directions). The Levi-Civita symbol is defined as $\epsilon^{txyz\r} = - \epsilon_{txyz\r} = 1$. 
$f_{mn} = \partial_m a_n - \partial_n a_m $ corresponds to the field strength of the fluctuations of the gauge field (\ref{AA}). 

The action \eqref{eq.generalmeson} consists of a Maxwell term plus a topological term\footnote{Note that by studying the topological term we may be able to find instabilities in probe brane systems which exhibit a spatially modulated phase. According to \cite{Nakamura:2009tf} the addition of a topological term to the Maxwell action may induce such an instability and using this notation we can tune the coupling as an external parameter. See also \cite{Ky} for this type of instability in the context of the chiral magnetic spiral of the Sakai-Sugimonto model.}. The second term may have some formal relation to a higher derivative 
correction~\cite{Myers:2010pk}. The OSM $s_{mn}$ naturally raises indices and the density $\sqrt{-s}$ is compensated by the running gauge coupling $g_{5}^2$. $Q_m$ and $s_{mn}$ encode the background gauge field effect such as of a finite density or background electromagnetic field. The topological term appears only when there are two or more non-vanishing $\theta^{mn}$ elements with all different indices, \eqref{eq.gm2}.  Although the use of the OSM and $\theta$ has appeared previously in the literature \cite{Albash:2007bq}, this is the first time, to our knowledge, that the effective action \eqref{eq.generalmeson} is written explicitly, emphasising the topological nature of the second term (in \cite{Gibbons:2000xe} the effective actions for scalars, spinors and gravitons were discussed and in \cite{Landsteiner:2007bd} the importance of the effective metric in the absence of probe branes was stressed.).

 Note that this effective action is not complete since there are also coupling terms between gauge and scalar fields (fluctuations transverse to the $D7$-brane) in general. 
For a fully general discussion of gauge fluctuations, we have to consider scalar fields dynamics and their coupling to gauge fields together. 
However, in order to highlight our proposal of using the OSM in the simplest setup, 
we choose to work using examples that can be described effectively only by (\ref{eq.generalmeson}) in the following sections. 
For this purpose we will restrict ourselves to the spatial component of the gauge field, $a_i$ ($i=1,2,3$), at zero spatial momentum in the background of $\theta^{ui} = 0$. See the appendix for details\footnote{We thank the referee for drawing our attention to this subtlety.}.

\subsection{The membrane paradigm in the presence of background fields (I) \hspace{3cm} \\ 
(A generalization to non-diagonal spatial metrics)}\label{sec.mempar1}

We shall now derive the membrane paradigm for our action \eqref{eq.generalmeson} in the case in which $s_{mn}$ is static, and therefore there are no off-diagonal temporal components. The canonical momentum to $a_i$ $(i=x,y,z)$ at fixed radial variable $\r$ yields

\be
  {\cal J}^i(\r) = - \frac{\cal{N}'}{g_{\mathrm{5}}^2} \sqrt{-s}\, f^{\r i}  
  - \frac{\mathcal{N'}}{2} \epsilon^{mnp\r i}f_{mn}Q_p \ , \label{GeneralJ}
\ee
The expectation value ($j^\mu$) of the conserved current 
in the boundary field theory is identified with 
$-{\cal J}^i(\r \rightarrow 0)$ and the conductivity tensor ($\sigma^{ij}$) can be written as\footnote{Note that there is
a minus sign in the second term, which is due to a coordinate inversion $u \sim 1/r^2$.
This sign will be compensated by a minus sign in (\ref{EF1}). Thus the final result does not depend on the
coordinate choice. }
\be
  j^i(k^\mu) = 
   - {\cal J}^i(\r \rightarrow 0)(k_\mu) \equiv \sigma^{ij}(k_\mu) f_{jt}(\r \rightarrow 0) \, .
\ee

Let us consider perturbing the system with an infinitesimal constant electric field in the $x$-direction, $f_{xt}={\cE}_x$, and compare the linear response expression with the holographic current. Although the holographic expression is taken at the $AdS$ boundary, in the hydrodynamic limit (corresponding to the DC response), the current can be shown to be invariant under holographic flow for abelian fields
\be
\partial_u {\cal J}^i(u)|_{\omega\rightarrow 0}=0 \, , \label{check1}
\ee
and thus the current can be calculated at any position in the radial direction. 
Just as discussed for the original membrane paradigm, if the OSM exhibits a horizon we can use the fact that the gauge field strength should depend only on the ingoing Eddington-Finkelstein coordinate to make a constraint between the $f_{\r i}$ and $f_{ti}$ components of the field strength $f$
\be
f_{ui}= - \sqrt{\frac{s_{uu}}{-s_{tt}}}f_{ti} \, . \label{EF1}
\ee

We can thus use this to write $f^{ui}$ in terms of the electric field fluctuations as $f^{ui}=\frac{s^{ix}}{\sqrt{-s_{uu}s_{tt}}}\cE_x$. This leads to the holographic expression for the conductivity
\be
{\cal J}^i=-\frac{\mathcal{N}'}{g_{5}^2}\sqrt{\frac{s}{s_{tt}s_{uu}}}s^{ix}\cE_x + \mathcal{N}'\epsilon^{txiju}\cE_x Q_j\Bigg|_{u\rightarrow 1} \, .
\ee

Comparing this with the QFT expression from linear response theory for the conductivity, we arrive at the following expressions for the direct and Hall conductivities in the presence of background fields written in terms of the open string degrees of freedom
\beqa
\sigma^{xx} &=& \left.\frac{\mathcal{N}'}{g_{5}^2}\sqrt{\frac{s}{s_{\r\r}s_{tt}}} s^{xx}\right|_{\r\rightarrow 1} \label{s1} \, ,\\
\sigma^{yx}& =& \left.\frac{\mathcal{N}'}{g_{5}^2}\sqrt{\frac{s}{s_{\r\r}s_{tt}}} s^{yx}
                  - \mathcal{N}'Q_z\right|_{\r\rightarrow 1} \label{s2}  \, ,\\
\sigma^{zx} &=& \left.\frac{\mathcal{N}'}{g_{5}^2}\sqrt{\frac{s}{s_{\r\r}s_{tt}}} s^{zx}
   + \mathcal{N}' Q_y \right|_{\r\rightarrow 1} \label{s3}\, .
\eeqa

\subsection{Conductivity with an external magnetic field at finite baryon density}

As an example, we now show how this result recovers the known results for conductivities in the presence of finite magnetic field and baryon density on the probe brane. In this case we turn on a gauge potential of the form
\be
  \tA =  \tA_t(u)\,\dd t + \tB_z x\, \dd y + \tB_x y \, \dd z   \, , \label{A}
\ee
where we could have set either $\tB_x=0$ or $\tB_z=0$  using $O(3)$ symmetry, but here we keep both components to show how the result in the previous section gives the right answer straightforwardly, even when we do not take the simpler ansatz. This background gauge field leads to an open string metric with non-diagonal terms in the $x$ and $z$ directions. 
\begin{eqnarray}
\mathrm{d} s^2 &=& s_{mn} \mathrm{d}x^m \mathrm{d}x^n\\
&= & g_{tt}{\cal G}^2 \dd t^2+\frac{\left(\tB_x \dd z-\tB_z \dd x\right)^2}{g_{xx}}+g_{xx} \left(\dd x^2+\kappa\, \dd y^2+ \dd z^2\right)+g_{\r\r}{\cal G}^2 \dd \r^2+g_{\Omega\Omega}\dd\Omega_3^2 \, ,\nonumber
\end{eqnarray}
where
\be
{\cal G}^2 \equiv \frac{ \kappa\, g_{\Omega\Omega}^3{g_{xx}^3}}{\hat{n}_q^2+\kappa\, g_{\Omega\Omega}^3{g_{xx}^3}}\, , \quad
 \kappa \equiv 1+\frac{{\tB_x^2}}{{g_{xx}^2}}+
\frac{{\tB_z^2}}{{g_{xx}^2}} \, .
\ee
The running coupling is given by $g_{5}^2=\kappa^{1/2}{\cal G}$, and the topological term receives contributions from
\be
Q = \frac{\hat{n}_q \left( \tB_x \dd x+ \tB_z \dd z \right)}{g_{xx}^2\kappa}\, .
\ee

Thus, by \eqref{s1}-\eqref{s3} we obtain
\beqa
  \sigma^{xx} &=& \frac{N_f N_c T}{4 \pi }\sqrt{\left(1-\psi_s^2\right)^3  + \frac{\tn_q^2}{1+\tb_x^2+\tb_z^2}    }\frac{1+\tb_x^2}{\sqrt{1+\tb_x^2+\tb_z^2}} \ , \label{sxx1} \\
   \sigma^{yx} &=& -\frac{N_fN_c T}{4\pi} \frac{\tn_q \tb_z}{1+\tb_x^2+\tb_z^2}  \ , \qquad \sigma^{zx} = \frac{\tb_x \tb_z}{ 1+ \tb_x^2} \sigma^{xx} \ .
\eeqa
This agrees with the macroscopic conductivity, (\ref{sxx0}) and (\ref{szx0}) in the limit $E_x \rightarrow 0$. The conductivity (\ref{sxx1}) reduces to (\ref{Jx1}) for small electric field and zero magnetic field.  

Notice that this is a non-trivial result. By using the open-string membrane paradigm we have trivially obtained the DC limit of the retarded current-current correlator without having to explicitly study the $\r$ dependence of the fluctuations. The full calculation which would otherwise have to be undertaken and was performed explicitly in \cite{Tarrio:2009vk} is summarized in the following steps: (1) study the behaviour of the (coupled) equations of motion for the gauge field fluctuations at singular points (horizon and boundary), imposing ingoing-wave boundary condition at the horizon and normalizing the value of the fluctuation at the boundary \cite{Son:2002sd}. (2) Decouple the equations of motion by performing a  change of basis, from cartesian to circular polarization (around the axis defined by the magnetic field). (3) Solve the differential equations in the hydrodynamic approximation and find the ratio between the normalizable and non-normalizable modes at the boundary to obtain the retarded Green's function, changing the basis back to the cartesian polarisation. (4) The linear-in-frequency, antihermitian part of the retarded Green's function corresponds to the DC conductivity tensor. Clearly using the membrane paradigm recipe we have a much simpler algorithm for finding these transport coefficients.

%%% STARTS ELECTRIC CASE

\section{The singular shell as a horizon and the electric membrane paradigm}\label{sec.sshell}

The phenomenon of emergent horizons in probe brane physics is not a new one. Indeed the simplest example is that of the rotating $D1$-brane or fundamental string (see \cite{Das:2010yw} for examples). Taking the brane to lie in the radial direction of $AdS$ and spinning it in one of the angular directions of the $S^5$ leads to two interesting properties. The first is that the induced metric on the brane reduces to an $AdS_2$ black hole with a temperature proportional to the brane's angular velocity -- a world-volume horizon emerges on the open string. The second important point is that there is a singularity at the centre of the $AdS$ space. This is expected due to the probe approximation. An external force is adding energy to the string to accelerate it. The string interacts with the background geometry only through the induced metric and does not backreact. The signal that the energy being pumped into the string has nowhere to go is that a singularity emerges inside the horizon. Precisely the same thing happens in the case of an electric field on a $Dq$-brane as we will see below. In fact, the rotating brane and the brane in a background electric field are simply related by a T-duality, but whereas the T-dual picture has been discussed in \cite{Filev:2008wj, Filev:2009xp} , we would like to show here how the horizon emerges naturally in the open-string picture.

Having recovered all previous results for conductivities from linear response theory, we would like to examine the more interesting cases in which the open string metric has a horizon which does not coincide with the horizon of the background spacetime. Generically this happens when there is a background electric field turned on on the probe brane. In this section we will first discuss the interesting properties of this horizon in a variety of situations before returning to the calculation of the conductivity in the presence of a macroscopic electric field, defined on what will be termed the \emph{electric membrane}.

The open string metric, defined in equation \eqref{eq.sab} in the case of finite electric field, current and baryon density is
\begin{eqnarray}\label{eq.effectivemetric}
\mathrm{d} s^2 = s_{mn} \mathrm{d}x^m \mathrm{d}x^n &= & \left( \tE_x^2+g_{tt} g_{xx} \right) \left( \frac{ \mathrm{d} \bar t^2 }{g_{xx}} + \frac{ \mathrm{d} \bar x^2 }{g_{tt}}\right)+ \frac{1}{g_{\r\r} } \left( \tA_t' \mathrm{d} \bar t + \tA_x' \mathrm{d} \bar x \right)^2\\
&& + \left( g_{\r\r} + \frac{g_{xx} \tA_t'^2+g_{tt} \tA_x'^2}{\tE_x^2+g_{tt}g_{xx}} \right) \mathrm{d} \r^2 + g_{xx} \mathrm{d}y^2 + g_{xx} \mathrm{d} z^2 + g_{\Omega_3\Omega_3}\mathrm{d}\Omega_3^2 \ , \nonumber
\end{eqnarray}
where the first and second components of the metric are the $\bar{t}$ and $\bar{x}$ directions defined by
\be
\mathrm{d} \bar t = \mathrm{d} t + \frac{\tE_x\, \tA_x'}{\tE_x^2+g_{tt}g_{xx}} \mathrm{d} \r\ , \qquad \mathrm{d} \bar x = \mathrm{d} x - \frac{\tE_x\, \tA_t'}{\tE_x^2+g_{tt}g_{xx}} \mathrm{d} \r \, .
\ee 
and the metric components $g_{\mu\nu}$ are those of the closed string metric. 
From now on we will work with the barred quantities, dropping the bar for simplicity.
The directions in this metric are the four Minkowski directions, 
the radial direction of the AdS space, $\r$, and the $S^3$ which is wrapped by the D7-brane.

We can now rewrite the effective metric \eqref{eq.effectivemetric} with the conserved charges 
and the position of the singular shell using the equations of motion for the gauge fields. 
More explicitly the metric components in \eqref{eq.effectivemetric} are given by\footnote{These are 
for $\psi_s = 0$. For $\psi_s \ne 0$ the expressions are more complicated, but 
the main physical structure of the metric is the same as the $\psi_s = 0$ case. }
\beqa
  && s_{tt} =  - \D_+\frac{\pi^2}{\r}\frac{P_1}{P_2} \ , \qquad
     s_{\r\r} = \frac{1}{\D_-} \frac{1+\te_x^2}{4 \r^2}\frac{1}{P_2}\ ,  \\
  && s_{xx} =  \frac{\pi^2(1+\te_x^2)(1+\r^3\tn_q^2)}{\r}\frac{P_1}{P_2} \ , \\
  && s_{tx} = - \pi^2 \r^2 \tn_q \te_x \sqrt{(1+\te_x^2)(\tn_q^2+(1+\te_x^2)^{3/2})}\frac{P_1}{P_2} \ ,
\eeqa
where
\beqa
   \D_+ &=&  (1-\r^2)(1+\te_x^2) - \r^3 \te_x^2 (\tn_q^2 + (1+\te_x^2)^{3/2}) \ , \\
   \D_- &=&  1-\sqrt{1+\te_x^2}\ \r \ , \\
    P_1 &=& 1 + \sqrt{1+\te_x^2} \ \r \ , \\
    P_2 &=&  \left(1+\sqrt{1+\te_x^2} + \te_x^2 \r^2\right)\left(1+\te_x^2\right) 
  + \tn_q^2 \r^3\left(1+\sqrt{1+\te_x^2}\ \r\right) \ .
\eeqa
Note that $P_1 >0$ and $P_2 >0$ while $\D_-$ and $\D_+$ change sign.
Therefore $s_{xx} > 0$ and $s_{tx} < 0$ do not change sign, 
while $s_{tt} < 0 $ and $s_{\r\r} >0$ for small $\r$ and change sign at
some larger value of the radial variable (see figure \ref{fig:gkerr} for an example).

\begin{figure}[]
\centering
  {\includegraphics[scale=0.6]{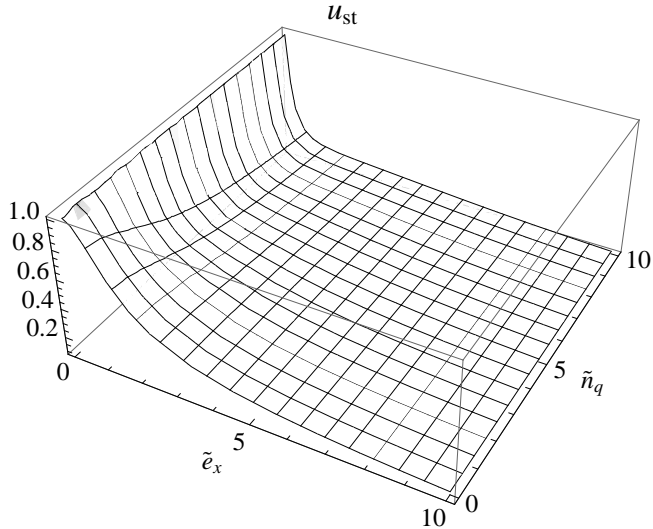}}
\caption{\em \label{fig:Sshell} The location of the stationary shell ($u_{st}$). 
For a given $E_x$ the stationary shell moves away from the horizon as $n_q$ increase.
}\end{figure}

The structure of the effective metric \eqref{eq.effectivemetric} has similarities to the Kerr black hole. 
Since the metric is stationary we have an event horizon at $s^{\r\r}=0$  (equivalent to $ \D_- = 0)$. That is 
\begin{eqnarray}
  s^{\r\r}(u_s)=0\ (\mathrm{event\ horizon}) \quad  \mathrm{where}
  \quad  u_s &=& \frac{1}{\sqrt{1+\te_x^2}}\ (\mathrm{singular\ shell}) \ ,
\end{eqnarray}
where now the singular shell plays
the role of an event horizon. Furthermore the hypersurface at $u=u_{st}$ such that $s_{tt}(u_{st})=0$ 
(or equivalently $\Delta_+(u_{st}) = 0$) plays the role of 
a \emph{stationary shell}.  In figure \ref{fig:Sshell} we present a plot of the location of the stationary shell as a function of the electric field strength and baryon density.
At $n_q = 0$ the stationary shell and singular shell are degenerate.

Between the event horizon and the stationary shell there will be an {\it ergoregion} similar to the Kerr black hole's ergosphere.  In the case of zero density, $A_t = 0$, the metric is completely diagonalized and the event horizon
and the stationary shell coincide.  Having found the structure of the open string metric we can now define an open string membrane paradigm on our open string horizon.

\subsection{The open string black hole temperature at zero baryon density}

We restrict for now to the case of zero baryon density, in which the metric \eqref{eq.effectivemetric} is diagonal and therefore the calculations are simpler. We will reobtain the results of this section in the $n_q\to0$ limit case of the most general treatment given in the next section. For the case at hand, the \emph{near-horizon} metric is given by:
\begin{eqnarray}
 \frac{\mathrm{d}s^2|_{u\rightarrow u_s}}{\pi^2L^2}&=&-2 T^2 \left(1+\tilde{e}_x^2\right) \r\, \mathrm{d} t^2+T^2\sqrt{1+\tilde{e}_x^2}\left( \mathrm{d} y^2+ \mathrm{d} z^2\right) 
\nonumber\\
&&+\frac{\left(1+\tilde{e}_x^2\right)^\frac{3}{2}}{\left(2+3\tilde{e}_x^2\right)}\left(\frac{2 T^2 \mathrm{d} x^2}{\mathcal{H}}+\frac{L^2\mathrm{d} \r^2}{2\left(1+\r\mathcal{H}\right)}\right)+L^2\left(1-\psi_s^2\right) \mathrm{d} \Omega_3^2 \, ,
\end{eqnarray}
with
\be
\mathcal{H}=\sqrt{\frac{1}{(1-\psi_s^2)}\left(1-\frac{\left(4+3\tilde{e}_x^2\right)}{\left(2+3\tilde{e}_x^2\right)^2}\psi_s^2\right)}\, ,
\ee 
It is clear that, after scaling $x$, this is a Rindler space times $\mathbf{R^3}\times S^3$. We can calculate the associated temperature 

\be\label{eq.temp1}
T_{\mathrm{eff}}=\frac{ T\sqrt{2+3\tilde{e}_x^2}}{2 \left(1+\tilde{e}_x^2\right)^\frac{1}{4}}\sqrt{1+\mathcal{H}}\, .
\ee
The final result is a measure of the effective temperature felt by the fundamental matter in the presence of a bath of finite temperature adjoint matter and a finite electric field. Essentially, the fundamental degrees of freedom feel not only the adjoint matter, but also the newly pair created fundamental particles of opposite charge moving in the opposite direction.

The dependence on the mass is captured through $\psi_s=\psi_s(T,E_x,m_q)$. It is worth noting that this effective temperature depends on the mass of the fundamental matter, the external electric field we have turned on and the temperature associated to the \emph{original} black hole in our $D3$-brane background. The open string degrees of freedom effectively feel a larger temperature than the one given by the background black hole. Such effects have been discussed in \cite{Liu:2006nn} in the presence of the trailing string.
For $\te_x (\sim E_x/T^2) \gg 1$, the effective temperature is    
\begin{eqnarray}
  T_{\mathrm{eff}} \rightarrow \sqrt{\frac{3}{2\pi\sqrt{\lambda} (1-\psi_s^2)}} E_x\ ,
\end{eqnarray}
while, for $\te_x \ll 1$, $T_{eff} \rightarrow T$.

There is also an interesting divergence which occurs in equation \eqref{eq.temp1} at the threshold electric field needed for pair creation. At this value of the electric field, $\psi_s=1$ and we can see that the temperature diverges. There are several notable factors here. The first is that this is not a stable solution. There is a phase transition for some $\psi_s\lesssim 1$, after which the system is described by embeddings that do not reach the singular shell. The second is that this is at a quantum critical point. One might expect that at the quantum critical point, as correlation lengths diverge, that thermal fluctuations will become increasingly important and thus the effective temperature should diverge in this region. One way to see this divergence may also be that it is due to the enthalpy of dissociation as we move from a bound phase to a dissociated phase of mesons.

\subsection{Finite baryon density and the drift of the electric membrane}

In the previous section we set the baryon density to zero which simplified the analysis of the effective metric. In this section we turn the baryon density back on in order to study the new phenomena which arise in this case.

\begin{figure}[]
\centering
\includegraphics[scale=0.8]{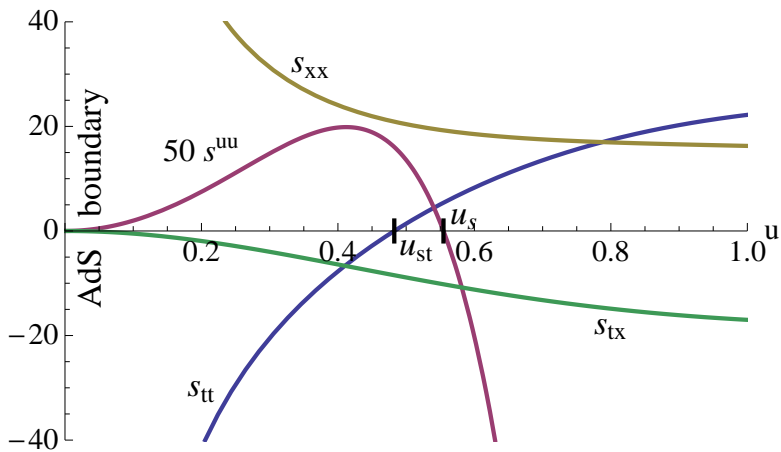}
\caption{\em \label{fig:gkerr}
Radial dependence of the induced metric components $s_{tt} $ (blue line), $s_{xx}$ (yellow line) $s^{\r\r}$ (red line, scaled to fit on the same plot) and $s_{tx}$ (green line) for $T=1/(\pi L)$, $\te_x=1.5$, $\tn_q=2$ and $\psi_s=0$ (corresponding to massless fundamental matter). The metric components asymptote to the $AdS$ values in the $u\rightarrow 0$ limit.
}
\end{figure}

The position of the singular shell, $\r_s$, is defined by the position at which the DBI action has to be regularized by the introduction of a current. However, looking at the effective metric \eqref{eq.effectivemetric}, it is clear that at finite baryon density the component $s_{tt}$ does not vanish at this position (in contrast to the zero baryon density case). In figure \ref{fig:gkerr} we plot the radial dependence of the induced metric components $s_{tt}$, $s_{xx}$, $s^{\r\r}$ and $s_{tx}$ showing that there are now two radii of interest, as discussed in the previous section.

In this figure we observe the presence of two relevant surfaces. One of them is the singular shell discussed before (at $\r_{s}=1/\sqrt{1+\tilde{e}_x^2}$, independent of baryon density) which represents the position of the event horizon of the induced metric. At the singular shell  $s^{\r\r}$ goes to zero linearly as $(\r_s-\r)$.

The second relevant surface is given by the
stationary shell, $\r_{st}$,  at which
$s_{tt}$ changes sign. Due to this change in the behaviour of
$s_{tt}$, objects between the stationary shell and the singular shell
are dragged along in the $x$-direction. This can be understood as the
dragging produced by the charged particles (since we are working with
finite baryon density) when accelerated by the electric field in the
$x$-direction. This effect is very similar to the rotational
frame-dragging which occurs in Kerr black holes, thus the name \emph{ergoregion} (\emph{nb.} not the ergosphere as the black hole is planar. The relation between this geometry and the Kerr black hole is thus a local one and not a global one).

The radius at which the stationary shell sits  relative to the
singular shell is dependent on the baryon density and the mass of the
fundamental matter. When the baryon density vanishes the stationary
shell merges with the singular shell and effectively disappears. The curvature invariants for both the original induced metric and the open string metric are everywhere finite. We can find the rate of frame dragging by studying the Killing vectors of the open string metric.

Consider the vector
\be\label{eq.killing}
\xi = \cosh\eta\, \partial_t + \sinh\eta\, \partial_x \, ,
\ee
where
\be
\sinh\eta  =  \frac{\tn_q \te_x}{\sqrt{\tn_q^2 + \left(  1+\te_x^2 \right)^{\frac{5}{2}} \left( 1-\psi_s^2 \right)^3 }} \, .
\ee
This is a Killing vector such that the norm squared, $\xi\cdot \xi$, vanishes at the singular shell and coincides with the behaviour of the $AdS$ metric component $g_{tt}$ at the boundary, giving the most natural normalization for the timelike Killing vector in the $AdS$/CFT context. It can be shown that this norm is always negative, such that this is timelike. Notice that when $\te_x\to0$ or $\td\to 0$ then $\xi=\partial_t$, recovering the Killing vector of Schwarzschild-$AdS$. With this definition of $\xi$ we can find the speed at which the singular shell moves, on a similar footing to the determination of the angular speed of a rotating black hole. This speed is given by
\be\label{eq.vs}
v_{s}^2 = \tanh^2 \eta = \frac{\tn_q^2 \te_x^2}{\left( 1+\te_x^2 \right)  \left( \tn_q^2 + \left( 1+\te_x^2 \right)^{3/2}\left( 1-\psi_s^2  \right)^3 \right)} \leq 1 \, ,
\ee
with the velocity approaching $1$ in the $\td\to\infty$ limit or in the simultaneous one $\te_x \to \infty$ and $\psi_s\to 1$. Rewriting $\te_x$ and $\tn_q$ in term of the $E$ and $n_q$ variables we can take the extremal limit, and the speed takes the form
\be
v_s^2 = \frac{n_q^2}{ n_q^2+ \frac{N_f^2 N_c^2}{8\pi^3\sqrt{\lambda}}E_x^3  \left( 1-\psi_s^2  \right)^3} \, .
\ee

Equation \eqref{eq.vs} describes a net velocity of the charged fundamental matter carriers driven by the electric field, and its product with the electric density of the quarks, $n_q$, can be understood as a current 
\begin{eqnarray}
  J_{n_q} \equiv n_q v_s = \frac{N_f N_c T}{4\pi} \frac{\tn_q^2}{1+\te_x^2} 
  \frac{1}{\sqrt{\sqrt{1+\te_x^2}(1-\psi_s^2)^3 + \frac{\tn_q^2}{1+\te_x^2}}} E_x \, , 
\end{eqnarray}
which is the contribution to the current due to the existence of charged matter dragged by the external electric field. 
This description becomes more accurate in the limit where the pair creation contribution is negligible and a good quasi-particle description is expected (for example, large mass and/or high density limit). In this limit
\begin{eqnarray}
  J_{n_q} \rightarrow \frac{N_f N_c T}{4\pi} \sqrt{\frac{\tn_q^2}{1+\te_x^2} } E_x \ ,
\end{eqnarray}
which agrees with (\ref{Jx1}). 
We see how the OSM description naturally incorporates this by describing a membrane moving at precisely $v_s$ at the horizon.

There is a second contribution to the current due to thermal pair creation. As explained in section \ref{sec.metallic}, this contribution will not contribute to the net momentum flow, since particles and antiparticles will be moving in opposite directions, but due to the charge inversion a net current will be present. We will use the membrane paradigm to find the total conductivity of the system, including the effects of pair creation as well as the net flux of charges from the finite background density.

\subsection{The open string black hole temperature at finite baryon density}

As the singular shell is a Killing horizon, this allows us to calculate the surface gravity $\kappa$ by
\be
\kappa^2 = -\frac{1}{2} (D_\mu \xi_\nu)(D^\mu \xi^\nu)  \, ,
\ee
from here we can read the temperature of the associated black hole as
\be\label{eq.teff}
T_{\mathrm{eff}}^2 = \frac{\kappa^2}{4\pi^2} =\frac{T^2}{4}\frac{\left( W_1 + \sqrt{ W_1^2 + W_2^2} \right) }{ \tn_q^2+
\left(1+\tilde{e}_x^2\right)^{5/2}
\left(1-\psi_s^2\right)^3 }\ ,
\ee
where
\beqa
W_1 &\equiv& 2 \tn_q^2 \sqrt{1+\tilde{e}_x^2}+
\left(1+\tilde{e}_x^2\right)^2 \left(2+3\tilde{e}_x^2\right)
\left(1-\psi_s^2 \right)^3 \ ,\\
W_2 &\equiv& 3 
\tilde{e}_x \psi_s \left(1+\tilde{e}_x^2\right)^{5/2} \left(1-\psi_s^2\right)^{5/2}\ .
\eeqa
In the zero density limit this result agrees with the previous calculation of the induced temperature given by the behaviour of the Rindler horizon \eqref{eq.temp1} and in the infinite density limit  $T_{\mathrm{eff}} \rightarrow  T (1+ \te_x^2 )^{1/4}$.
Notice that finite density can regulate the divergence at zero density and $\psi_s = 1$.   
As $T \rightarrow \infty $ or $E \rightarrow 0$, $T_\mathrm{eff} \rightarrow T$, while as $T \rightarrow 0$ or $E \rightarrow \infty$ , $T_{\mathrm{eff}} \rightarrow 
\sqrt{ \frac{3 E}{2\pi\sqrt{\lambda}}\left(1+\frac{1}{\sqrt{1-\psi_s^2}}\right)}$.

There are some interesting things to note about this effective temperature. Such effective temperatures have been discussed previously (\emph{e.g.}, \cite{Liu:2006nn}) in the context of trailing strings \cite{Herzog:2006gh, CasalderreySolana:2009rm}, 
where a quark moving at equilibrium in a background plasma feels an effective temperature caused by a boost of the stress energy tensor in its frame. This leads to an effective temperature of the form $T_{\mathrm{eff}}=(1-v^2)^{-\frac{1}{4}}T$ therefore diverging as the velocity approaches the speed of light. In the current case it is harder to see the velocity because although we have a steady current, the charge carriers are constantly being pair created, accelerated and then annihilated. As discussed in the previous section the velocity can be seen in the case of finite baryon density, due to the net momentum flow of surplus charge carriers. This velocity only goes to one as the electric field strength diverges, and indeed in that case the temperature can also be seen to diverge.

\subsection{A new electric membrane paradigm}

Now we will return to the membrane paradigm calculations that were discussed in section \ref{sec.mempar1} showing how, with the electric membrane as a true horizon on the $D7$-brane from the open string point of view, we can recover the electric conductivity in the presence of a macroscopic electric field.

As a warmup let us consider the zero density case, in which the open string metric is static and we can use the result \eqref{s1} with a horizon determined by the singular shell at $\r_s = 1/\sqrt{1+\te_x^2}$. We have
\begin{eqnarray} \label{eq.zeronbconduc}
  \sigma^{xx}  &=& \left.\frac{\mathcal{N}'}{g_{5}^2}\sqrt{\frac{s}{s_{\r\r}s_{tt}}} s^{xx}\right|_{\r=1/\sqrt{1+\te_x^2}} \nonumber \\
  &=& \frac{N_f N_c T}{4 \pi }\frac{(2+3 \te_x^2)\sqrt{\sqrt{1+\te_x^2}(1-\psi_s^2)^3}}{2(1+\te_x^2)}\ ,
  \end{eqnarray}
  But this expression can be written as the derivative with respect to the electric field of a simpler quantity, given by:
  \begin{equation}
 \sigma^{xx}=   \frac{N_f N_c T}{4 \pi } \partial_{\te_x} \left[\sqrt{\sqrt{1+\te_x^2} \left(1-\psi_s^2 \right)^{3}} \, \te_x \right]\ ,
  \end{equation}
  which is simply:
\begin{equation}
 \sigma^{xx} =   \frac{N_f N_c T}{4 \pi } \partial_{\te_x} \big[\tJ_x(\te_x)\big]  \label{sE}   \, .
\end{equation}
This $\tJ_x$ is that obtained in \eqref{Jx}. This result is understood as follows. The membrane paradigm is based on linear response theory, so a small source, $\delta E_x (\equiv \cE_x)$, should be assumed.  
However we already have a large background $E_x$ and corresponding $J_x$, 
which are encoded in the metric. 
Thus, the linear response will describe the response to a small $\cE_x$ on top of the background $E_x$,  \emph{i.e.}
\begin{eqnarray}\label{eq.lrtcurrent}
  J_x(E_x + \cE_x) - J_x(E_x) &\sim&   \partial_{E} \left(J_x(E)\right) \cE_x \nonumber \\
      &=& \frac{N_f N_c T^2}{4\pi} \partial_{\te} \left(\tJ_x(\te)\right) \cE_x  \nonumber \\
      &=& \sigma^{xx} \cE_x \ ,
\end{eqnarray}
which is thus confirmed in \eqref{sE}. This result was obtained with the evaluation at the OSM horizon $\r_s = 1/\sqrt{1+\te_x^2}$ and thus confirms the expectation that we can define a membrane paradigm at this position. 

Recall that an explicit microscopic calculation in terms of the original variables $g_{mn}$ and $F_{mn}$ was performed in \cite{Mas:2009wf}. In this paper a small electric field in the $y$-direction was added to the background $E_x$, and the conductivity was found in linear response theory following the philosophy of equation \eqref{eq.lrtcurrent}. In this work it was found that the fluctuations had to satisfy a (regularity) boundary condition at the singular shell to recover the result \eqref{eq.zeronbconduc}, which was found macroscopically in \cite{O'Bannon:2007in}.  . The possible boundary conditions at the (closed string) horizon appeared strange for a regular black hole, since the modes at the horizon did not split into ingoing and outcoming waves, but were both ingoing waves, one of them with an extra damping factor. With the membrane paradigm based on the OSM we present in this paper, regularity is imposed at the singular shell, at which the modes decompose again as ingoing/outcoming waves (see section \ref{sec.QNMs} where the strange nature of these boundary conditions will become clear). 

Note that we are using a rescaled time. Since our $t$ is indeed $\bar{t}$, we have to be careful about interpreting the result, in principle. 
However, there is no problem in our case, since at the boundary $\r \rightarrow 0$, $\bar{t} \rightarrow t$. So our field theory interpretation will be the same for $\bar{t}$ and $t$.

\subsection{The membrane paradigm in the presence of background fields (II) \hspace{3cm} \\
(A generalization to  non-static, stationary metrics) } \label{sec.mempar2}

Here we will consider the new membrane paradigm not only for static metrics, but for\emph{non-static, stationary}  ones. There are thus off-diagonal components in the $t$-direction. Consider for simplicity a metric of the form
\be\label{eq.nondiagmet}
\dd s^2=s_{tt} \dd t^2+s_{xx} \dd x^2+2s_{tx}\dd t \dd x+s_{uu}\dd u^2 
+ s_{yy}\dd y^2 + s_{zz}\dd z^2 + s_{\Omega\Omega} \dd \Omega_3^2  \, ,
\ee
 By applying a generalized 
membrane paradigm to the finite density and electric field case (\ref{eq.effectivemetric}) 
we will confirm the conductivity obtained macroscopically, 
showing the usefulness of the OSM and a generalized membrane paradigm.

In order to calculate the conductivity ($\sigma^{ij}$) we must relate the current $j^{i}$ (conjugate momentum of the gauge field $a_i$, thus $\sim f^{ui}$)    
to the electric field strength ($f_{jt} = \cE_j$) on the $AdS$ boundary. In this example we set up the electric field in the $x$-direction.
The current is given by\footnote{We will omit the contribution from the topological term 
in (\ref{GeneralJ}), since the present discussion does not affect that part of the current.}
\be \label{J10}
{\cal J}^i(u)= -{\cal N}' \frac{\sqrt{-s}}{g^2_{5}(u)}f^{ui}(u) \, .
\ee
Because of the anisotropy in the spatial directions there are three currents we can write down (though in our case two will be identical)
\beqa
  && {\cal J}^y \sim f^{uy} =  s^{uu}s^{yy}f_{uy}  \ , \label{conductT}\\ 
  && {\cal J}^z \sim f^{uz} =  s^{uu}s^{zz}f_{uz}  \ , \label{conductTz}\\ 
  && {\cal J}^x \sim f^{ux} =  s^{uu}s^{xx}f_{ux} + s^{uu}s^{tx}f_{ut}   \ ,\label{conductL}   
\eeqa
where, in the last terms, we simply lowered the indices with the open string metric.

As before, in the hydrodynamic limit the current is independent of the radial direction, ($\partial_\r{\cal J}^i(\r)=0$), 
and the relations \eqref{conductT}-\eqref{conductL} hold at any radial position -- in particular, on the $AdS$ boundary where the field theory is defined and on the open string horizon, where the membrane paradigm will be used to relate the current to the electric field.
Following the membrane paradigm, this relation can be made through the constraint of the regularity of the field strength at the open string horizon which is equivalent to imposing incoming boundary conditions at this point. In order to do this we must generalize the Eddington-Finkelstein coordinates to the non-static metric of equation \eqref{eq.nondiagmet}\footnote{This is similar to the Kerr black hole case, where there are no radial null geodesics due to frame drag and one can define instead the \emph{principal null congruences} to define the Eddington-Finkelsetin coordinates.}.

By symmetry we can consider null geodesics at constant $y,z$ and $\Omega_3$, described
by $p^\mu \equiv \{\dot{x}, \dot{t}, \dot{u} \}$, where the dots  denote derivatives 
with respect to an affine parameter. Geodesic equations can be written down 
as three first integrals from two Killing vectors ($\partial_t , \partial_x$) and the null condition ($\dd s^2 = 0$):    
\beqa
&& p_t = s_{tt}\dot{t}+s_{tx}\dot{x}\  \, ,\quad  p_x = s_{tx}\dot{t}+s_{xx}\dot{x} \ \, , \\
&& \dd s^2 =s_{tt}\dot{t}^2+s_{xx}\dot{x}^2+2s_{tx}\dot{t}\dot{x}+s_{uu}\dot{u}^2=0 \, ,
\eeqa
whose solutions 
\begin{eqnarray}
  \dot{u} = \mp \sqrt{\frac{p_x^2 s_{tt} - 2 p_t p_x s_{tx} + p_t^2 s_{xx} }{ s_{uu} (s_{tx}^2 - s_{tt} s_{xx} )}  } \ , \quad
  \dot{t} = \frac{p_x s_{tx} - p_t s_{xx}}{s_{tx}^2 - s_{tt} s_{xx}} \ , \quad
  \dot{x} = \frac{p_t s_{tx} - p_x s_{tt}}{s_{tx}^2 - s_{tt} s_{xx}}\ ,
\end{eqnarray}
yield two principal null congruences corresponding to the two signs of
\beqa
\dd t & = & \pm \frac{\sqrt{s_{uu}} (s_{xx} - \alpha s_{tx} )}{\sqrt{s_{tx}^2-s_{tt}s_{xx}}\sqrt{s_{xx}-2\alpha s_{tx}+\alpha^2 s_{tt}}} \dd\r \equiv \pm \Lambda (u) \dd\r\, , \\
\dd x & = & \pm \frac{\sqrt{s_{uu}} (\alpha s_{tt} - s_{tx})}{\sqrt{s_{tx}^2-s_{tt}s_{xx}}\sqrt{s_{xx}-2\alpha s_{tx}+\alpha^2 s_{tt}}} \dd\r \equiv \pm \hat{\Lambda} (u) \dd\r\, ,
\eeqa
where $\alpha = p_x/p_t $ and the ingoing (outgoing) congruence corresponds to the plus (minus) sign. The sign 
can be identified by looking at the limit $s_{tx} \rightarrow 0$. We define an (ingoing) Eddington-Finkelstein coordinate system $\{v,\hat{x},u\}$ by 
\begin{eqnarray}
  \dd v = \dd t - \Lambda(u) \dd u \ , \quad \dd \hat{x} = \dd x - \hat{\Lambda}(u) \dd u  \ ,
\end{eqnarray}

Now that we have defined our regular Eddington-Finkelstein directions, 
we can constrain the gauge fields to depend on $x$, $t$ and $u$ only through these combinations at the horizon, 
which gives a condition 
\begin{eqnarray}
   (\partial_u + \Lambda \partial_t +  \hat{\Lambda} \partial_x ) a_\mu = 0  \ . \label{constraints}
\end{eqnarray}
Since we are interested in a homogeneous DC conductivity we assume that $a_i = a_i(t,u)$ so 
\eqref{constraints} yields 
\begin{eqnarray}
  f_{ui} = - \Lambda f_{ti} \label{Fuiti} \ .
\end{eqnarray}
This relation can be shown to be gauge invariant as follows.  
The regularity condition \eqref{constraints} of $f_{ti}$ and the Bianchi identity read
\begin{eqnarray}
  && \Lambda \partial_t f_{ti} = - \partial_u f_{ti} - \hat{\Lambda}\partial_i f_{ti} \ , \label{k1} \\
  && \partial_u f_{ti} = \partial_{t} f_{ui} + \partial_i f_{tu} \ ,\label{k2}
\end{eqnarray}
which imply that $\Lambda f_{ti}  +   f_{ui} = c $, where $c$ is a gauge-independent constant, if $f_{\mu\nu}$ is homogeneous.
From (\ref{Fuiti}) $c = 0$.  

With (\ref{Fuiti}) and $f_{ut} = - \hat{\Lambda} f_{xt}$, the currents \eqref{conductT}-\eqref{conductL} read 
\begin{eqnarray}
  && j^{i} = {\cal N}' \frac{e^{-\phi}}{g^2_{5}(r)} \sqrt{s} s^{uu} 
  (s^{ii} \Lambda - s^{ti}\hat{\Lambda}) \cE_i \ ,   \quad (i = x,y,z)  \ ,
\end{eqnarray}
For a metric like \eqref{eq.nondiagmet} with a Killing vector field given by \eqref{eq.killing}, with vanishing norm at the horizon, it is immediate to check that, at the horizon $-s_{tt}=2v_s s_{tx}+v_s^2 s_{xx}$, where $v_s=\xi^x/\xi^t$. Furthermore, as $s_{tt}s_{xx}-s_{tx}^2=0$ at this surface  too \cite{Pal:2010sx}, the components of the metric are related by
\begin{align}
& s_{tt}\Big|_{u_s} =  v_s^2 s_{xx}\Big|_{u_s}\, , \quad s_{tx}\Big|_{u_s} = - v_s s_{xx}\Big|_{u_s}\,, \\ 
\Rightarrow \quad &
  \Lambda\Big|_{u_s} = \frac{\sqrt{s_{uu}s_{xx}}}{\sqrt{s_{tx}^2-s_{tt}s_{xx}} }  \,, \qquad \hat{\Lambda}\Big|_{u_s} = - \Lambda \frac{s_{tx}}{s_{xx}}\Big|_{u_s}
\end{align}
where the $\alpha$-dependence of $\Lambda$ cancels at the horizon.
Therefore the conductivity reads 
\be\label{eq.transcond}
\sigma^{ii} =  \frac{{\cal N}'}{g_{5}^2} \frac{\sqrt{-s}}{\sqrt{s_{uu}}\sqrt{-s_{tt}s_{xx}+s_{tx}^2}}\frac{\sqrt{s_{xx}}}{s_{ii}}  \Bigg|_{u\to u_s}\ ,  \quad (i = x,y,z ) \,   .
\ee

Now let us apply this formula to the finite density and electric field case. 
In this case there is no topological term, but as we have seen the OSM is non-static, stationary with a non trivial $s_{tx}$ component. Plugging the metric components  in the conductivity formulae we recover the macroscopic result in \cite{O'Bannon:2007in}, solving for both the longitudinal and the transverse fluctuations of the gauge field along the direction of the electric field
\beqa
 \sigma^{xx} & = & \frac{N_f N_c T}{4\pi} \partial_{\te_x} \left[  \sqrt{\sqrt{1+\te_x^2} (1-\psi_s^2)^3 + \frac{\tn_q^2}{1+\te_x^2}} \te_x \right] \, , \label{eq.longcondEDis} \\
\sigma^{yy} &= & \sigma^{zz} = \frac{N_f N_c T}{4\pi} \sqrt{\sqrt{1+\te_x^2} (1-\psi_s^2)^3 + \frac{\tn_q^2}{1+\te_x^2}} \, . \label{eq.transcondEDis}
\eeqa
The microscopic calculation leading to \eqref{eq.transcondEDis} was performed explicitly in \cite{Mas:2009wf}, but the one leading to \eqref{eq.longcondEDis} is presented here for the first time. This result shows the power of the membrane paradigm, since the equations of motion governing the fluctuation along the $x$-direction are rather cumbersome.

\subsection{Microscopic excitations and the electric membrane} \label{sec.QNMs}

Having shown in the previous sections that we can recover the results for transport coefficients using the new membrane paradigm we can compare these results with the microscopic setup whereby the Kubo relation is explicitly used to calculate the conductivity \footnote{Note that the membrane paradigm method also uses the Kubo relation, but here it is an explicit numerical calculation whereas for the membrane paradigm it is implicit and analytic.}. This will be another check that the open string fluctuations really do see the membrane as a horizon. In \cite{Mas:2009wf} two of the authors of this paper performed this microscopic calculation in the closed string metric language and found that there was a unique set of boundary conditions for the fluctuations about the singular shell which recovered the known conductivity expression. However, we show here that in the open string language this calculation is much more transparent.

In order to calculate the conductivity we must use the Kubo relation
\begin{equation}
\sigma^{zz}=-\lim_{\omega\rightarrow 0}\frac{\mathrm{Im}\Pi^\perp}{\omega} \ ,
\end{equation}
where $\Pi^\perp$ is the transverse two point Green's function which in this case will correspond to the two point function of vector modes transverse to the direction of the electric field in the $x$-direction. Thus, using the usual holographic recipe we must calculate the equation of motion for the gauge field components $a_{y,z}(\r,t)$ and look at the boundary behaviour in the UV (since these two modes decouple and are described by the same equations we will focus only on $a_z$). The two point function is then found by looking at
\begin{equation}
\Pi^\perp=\lim_{\r\rightarrow 0}\frac{{a_z}'(\r)}{a_z(\r)}\ ,
\end{equation}
where, because we are linearising in fluctuations, $a_z(\r,t)=e^{i\omega t}a_z(\r)$ and where the field is normalised such that $a_z(\r\rightarrow \r_b)\rightarrow 1$ (see for example \cite{Mateos:2007yp} for an extensive treatment of such a calculation).

In the case of the original metric we can study the fluctuations and write the quadratic Lagrangian in the form
\begin{equation}
{\cal L}\sim \sqrt{-\gamma}\left(\mathrm{Tr}\left(\gamma^{-1}f\gamma^{-1}f\right) -
\frac{1}{2} \left(\mathrm{Tr} \left(\gamma^{-1}f \right)\right)^2\right)\ ,
\end{equation}
where here $f$ is the fluctuation of the gauge field about the background embedding. In the case of a finite electric field in the $x$-direction we can study the fluctuations of gauge fields in the transverse direction and get the following lagrangian (the functions $\mathbb{S}, \mathbb{T}$ and $\mathbb{U}$ can be found in \cite{Mas:2009wf})
\begin{equation}
{\cal L}\sim \sqrt{-\gamma}\left(\mathbb{S}(\r,E)\partial_t a_z (t,\r)^2+\mathbb{T}(\r,E)\partial_\r  a_z (t,\r)^2+\mathbb{U}(\r,E)\partial_t a_z (t,\r)\partial_\r a_z (t,\r)\right)\ .
\end{equation}
We see that there are cross terms between the different derivatives on $a_z(t,\r)$. From this Lagrangian the indices at the singular points can be calculated via the equations of motion. The singular shell being a singular point gives two indices which leads to two solutions to the gauge potential close to the singular shell, given by
\begin{eqnarray}
a_z(t,\r\sim \r_s)&=&e^{i\omega t}(\r-\r_s)^{\eta_{1,2}}\ ,\nonumber\\
\eta_1&=&0\, ,\quad \eta_2=-\frac{i (1+\tilde{e}_x^2)^{\frac{1}{4}}\omega}{\sqrt{2}\sqrt{2+3\tilde{e}_x^2}\pi T}\ .
\end{eqnarray}
These two indices don't look like the usual incoming wave boundary conditions that we are used to when studying quasinormal modes of black holes. By studying the $\tilde{e}_x \rightarrow 0$ limit and the hydrodynamic limit of the two point correlator of the vector current on the boundary, in \cite{Mas:2009wf} it was found that the zero index $\eta_1$  is the correct choice to recover the macroscopically calculated transport coefficients. In fact, this is precisely the incoming Eddington-Finkelstein boundary condition, but this fact is not clear in this coordinate system. We will show here that using the open string metric we are able to find the correct frame in which the incoming and outgoing Eddington-Finkelstein coordinates are transparent.

We can make the substitution $\gamma^{mn}=s^{mn}+ \theta^{mn}$ in the above set of equations and would find precisely the same result as above, but now with $\gamma$ replaced by $s$ in the effective action (up to an the overall prefactor  of $\gamma$). The cross term is seen to come from the non-diagonal form of $s$. We are free now to diagonalise the metric $s$ by a coordinate transformation (in the case of finite baryon density we can only remove the $\dd \r\, \dd t$ cross-terms and not the $\dd t \,\dd x$ ones). Now, writing the action with diagonalised metric we find a simpler form of the action for $a_z$ given by
\begin{eqnarray}\label{eq.fluclag}
-\frac{4 S_{\mathrm{eff}}}{{\cal N}'}&=&\int \dd^5x
\sqrt{-\gamma}s^{ac}s^{bd}f_{ab}f_{cd} \ ,\nonumber \\ 
&=& \int \dd^5 x 2\sqrt{-\gamma}\left(s^{zz}\left(\partial_t a_z(\r,t)\right)^2+s^{\r\r}\left(\partial_\r  
a_z(\r,t)\right)^2\right)\ .
\end{eqnarray}
Calculating the equation of motion for $a_z(\r,t)=e^{-i\omega t}a_z(\r)$ and performing a Frobenius expansion around the open string horizon we find that the indices are
\begin{equation}
a_z(\r)_{\r\rightarrow\r_s}\rightarrow(\r-\r_s)^{\pm i\frac{\omega}{4\pi T_{\mathrm{eff}}}}\ ,
\end{equation}
where $T_{\mathrm{eff}}$ is the effective temperature calculated in (\ref{eq.teff}). 
By combining the plane wave part of the ansatz, the boundary condition for $a_z$ at the singular shell is seen to be
\begin{equation}
a_z(\r,t)_{\r\rightarrow\r_s}=e^{-i \omega\left(t- \frac{1}{4\pi T_{\mathrm{eff}}} \log \left(\r-\r_s\right)\right)}\ .
\end{equation}
It is clear that using the open string metric in the diagonal coordinates we have found the natural Eddington-Finklestein ingoing coordinates. This analysis is a lot more opaque in the original  formulation.

Using these boundary conditions, the equation of motion coming from (\ref{eq.fluclag}) can be integrated to the $AdS$ boundary. Reading off the solution at the boundary and applying the Kubo relation a perfect fit with the macroscopic result for $\sigma^{zz}$ is found. This confirms the macroscopic calculation and also shows that the retarded Green's function is given by purely ingoing boundary conditions at the open string horizon.

The same calculation can be performed for the conductivity in the direction of the applied electric field and again there is a perfect fit between the microscopic and macroscopic solutions. Note that the transparency of this calculation will make the calculation of quasinormal modes for this system much simpler. We leave for future work.

\section{Conclusions}\label{sec.conclu}

In this work we have studied the effects of background electromagnetic fields on open string degrees of freedom on probe $Dq$-branes. Using the open string metric we have shown that the black hole membrane paradigm can be extended to complex cases involving non-diagonal metrics, non-static metrics and actions including topological terms for the fluctuating degrees of freedom. Moreover, the singular shell, discussed previously in many papers, has been shown to act as a true horizon for the open strings and allows for a membrane paradigm to be defined even when there is no horizon in the background spacetime.

From the field theory perspective we have been able to define the effective temperature felt by the fundamental matter in the presence of an electric field. It would be fascinating to compare this result with any similar field theory calculations. As noted, the effective temperature itself has some rather strange properties in the unstable regime and the effective temperature close to a quantum critical point is certainly worth investigating in more detail.

We noted in the introduction that the reason we obtain a finite conductivity in this work is because we are in the probe approximation. There would be no steady state in the back-reacted solution, but such a time-dependent solution may be tractable (see \cite{Karch:2010kt}) and would certainly be extremely interesting to study. 

The number of examples where gauge field configurations will induce a non-diagonal open string metric is huge, so the procedure presented in this paper has a huge potential to treat these. Some obvious directions for future work include the study of the entire fluctuation spectra (vectors, scalars and fermions) in terms of $s$ and $\theta$ and thus an investigation into the effects of the topological term on the gauge theory phenomenology, the extension of the current results to the case of the non-Abelian DBI action (where spontaneous condensation of vector fields can occur) or the inclusion of the Wess-Zumino term and its contribution to the topological term discussed in this paper. In particular the instabilities which may be driven by the topological term could be very important in a range of scenarios and although the methods described here will not give new solutions, they may highlight the areas that such phenomenology is likely to appear.

It is interesting to note that in the case of a large enough background field in an extremal background, because of the open string horizon, we have a theory which appears at first sight to be at zero temperature, and thus doesn't fall into the universality class of \cite{Iqbal:2008by} but through the dynamical generation of temperature will have the same universal value of $\frac{\eta}{s}=\frac{1}{4\pi}$.

The power and importance of the membrane paradigm in holography may yet provide more tools for studying universal properties of gauge theories and condensed matter systems. The extension of the membrane paradigm to more diverse situations is thus, we believe, an important topic for further investigation.

\acknowledgments

We would like to thank Johanna Erdmenger, Nick Evans, Veselin Filev, Hong Liu, Alexander Morisse, Andy O'Bannon, 
Miguel Paulos and Astrid Gebauer for helpful comments and discussions. 
J.T. is thankful to the Front of Galician-speaking Scientists for encouragement.

K.K. is grateful for the support of an STFC rolling grant.
J.S. would like to express his gratitude to the Marie Curie foundation and the Max Planck Institute for theoretical physics in Munich. J.T. is supported by the Netherlands Organization for ScientiÞc Research (NWO) under the FOM Foundation research program.

This work was initiated at the KITPC program "AdS/CFT and other novel approaches to hadron and heavy ion physics" in Beijing in Autumn 2010.

\appendix

\section{Coupling to scalar}

In general the coupling term to a scalar (which we denote by $\chi$) should be added to the action \eqref{eq.generalmeson}. Schematically it reads 
\begin{eqnarray}
  &&\mathcal{L} =  - \mathcal{N}' \frac{ \sqrt{-s} }{4\, g^2_{5}} s^{mp}s^{nq} f_{mn} f_{pq} 
   - \mathcal{N}' \frac{1}{8} \epsilon^{mnpqr} f_{mn}  f_{pq}Q_r  \nonumber \\ 
   && \qquad + (G_1 \theta^{mn} + G_2 \partial_u \theta^{mn})f_{mn} \chi \ , 
\end{eqnarray}
where $G_1$ and $G_2$ are functions of the background metrics and embedding profile, so are 
functions of $u$ only. We are interested in the case that spatial momentum vanishes, so the equations of motion for the $a_i(t,u)$ fields are
\begin{eqnarray}
  \partial_u \Pi^{ui} + \partial_t \Pi^{ti} = 0 \ ,
\end{eqnarray}
where the conjugate momentum $\Pi^{mn}$ is defined as
\begin{eqnarray}
  \Pi^{mn} = \frac{\partial \mathcal{L}}{\partial (\partial_m a_n)} \ .
\end{eqnarray}
Inserting a fourier mode decomposition and taking the zero frequency limit,we see that $\Pi^{ui}$ is conserved along $u$.  If $\theta^{ui} = 0$, as is the case in the examples we have considered, the conjugate momentum of interest to us reads
\begin{eqnarray}
  \Pi^{ui} = - \frac{\cal{N}'}{g_{\mathrm{5}}^2} \sqrt{-s}\, f^{\r i}  
  - \frac{\mathcal{N'}}{2} \epsilon^{mnp\r i}f_{mn}Q_p \equiv {\cal J}^i(\r) \ ,
\end{eqnarray}
which is our equation  \eqref{GeneralJ}. Thus, for this special case, we need not 
worry about scalar couplings and can safely ignore them in \eqref{eq.generalmeson} for simplicity.  
However, we emphasize that in general the coupling to the scalar should be 
taken into account.

\end{document}